\documentclass[apj,twocolumn]{openjournal}
\usepackage{amsmath, amssymb, amsfonts}
\usepackage{graphicx} % Required for inserting images
\usepackage{xcolor}
\usepackage{hyperref}
\newcommand{\dd}{\mathrm{d}}
\usepackage{appendix}

\hypersetup{
    colorlinks=true,
    citecolor=blue,
    linkcolor=blue,
    }

\begin{document}

\title{B--modes from galaxy cluster alignments in future surveys}
\author{Christos Georgiou$^\dagger$,
        Thomas Bakx,
        Juliard van Donkersgoed and
        Nora Elisa Chisari}
\email{$\dagger$ c.georgiou@uu.nl}
\affiliation{Institute for Theoretical Physics, Utrecht University, Princetonplein 5, 3584 CC, Utrecht, The Netherlands.}

%\date{July 2023}

\begin{abstract}
    Intrinsic alignment (IA) of source galaxies represents an important contaminant for upcoming cosmic shear surveys. In particular, it is expected on general grounds that IA contains a B--mode while the weak lensing signal does not. Thus, a detection of B--modes offers the possibility to study directly the IA signal of the sources. Galaxy clusters exhibit strong IA and are therefore a natural candidate to look for a B--mode signal. We forecast the signal--to--noise ratio (SNR) for B--modes from IA of galaxy clusters in the Vera C. Rubin Observatory Legacy Survey of Space and Time (LSST). We use a perturbative model for the IA multipoles based on the Effective Field Theory of Intrinsic Alignments (EFT of IA), which has recently been validated against N--body simulations. We forecast SNR $\approx 12$ and find that this detectability is not significantly impacted by different analysis choices. Lastly, we also apply our forecast to clusters in the redMaPPer SDSS and DESY1 samples. We find SNR $\approx 5$ and SNR $\approx 3$, respectively, suggesting a detection is within reach, provided accurate redshift information is available.
\end{abstract}

\maketitle

\section{Introduction}
\label{sec:Introduction}

Over the past decades, the $\Lambda$CDM model has emerged as the dominant paradigm for describing the history of the Universe and its various epochs across a wide range of redshifts with a small number of free parameters. On the other hand, given that several tensions in the measurements of these parameters between different probes within this model have reared their heads in recent years, it is no longer the case that $\Lambda$CDM is unchallenged \citep[see][for a recent review]{intertwined}. In addition, for future experiments such as the ESA \emph{Euclid} mission\footnote{\url{https://www.cosmos.esa.int/web/euclid}} \citep{Euclid} and the Vera C. Rubin Observatory Legacy Survey of Space and Time \citep[LSST\footnote{\url{https://www.lsst.org/}},][]{LSST}, we must face the challenge of controlling an ever--growing number of sources of systematics when analysing new datasets of cosmological probes. Crucially, our ability to continue improving our understanding of cosmology hinges on the ability to \textit{model} these systematics, rather than including them in inflated error budgets.

One probe of great importance is weak gravitational lensing \citep{WeinbergReview}: the coherent distortion of light from distant galaxies due to the gravitational pull of matter along the line--of--sight, also known as cosmic shear, which directly provides information on the statistics of the dark matter distribution \citep{bartelmann}. A difficulty in using weak lensing as a cosmology probe is the presence of several systematic errors that need to be controlled for an accurate measurement \citep[see, e.g.,][]{mandelbaum}. 

A prominent source of systematics for weak gravitational lensing is that of intrinsic alignments (IA): the correlation of the intrinsic, unlensed orientation of galaxies with large scale structure in its vicinity also causes coherent distortions, which are in principle degenerate with the lensing signal \citep{troxel,kiessling}. This effect has indeed been observed \citep[see, e.g.,][for recent measurements]{singh_boss,johnston,KiDS_LRG,samuroff_eboss}. 

While the alignment signal for luminous red galaxies (LRGs) is well understood, they only comprise a small portion of typical samples relevant for cosmic shear analyses. As such, the significance for cosmic shear analyses is as of yet unclear \citep{amon_desy,secco_desy, hsc_res, deskids}. Various theoretical models for the IA signal have been proposed \citep{catelan_ia,hirata_seljak,Schneider10,blazek_seljak,TATT,schmitz,vlah1,halo_model,Maion}. Among these, \citet{Schneider10} and \citet{halo_model} offer approaches based on the halo model, which requires making assumptions about the galaxy-halo connection and their IA. All other approaches are based on perturbation theory and they rely on fewer physical assumptions. However, they may not be able to extend to the smaller scales from which lensing derives much of its constraining power. The range of validity of these models is thus a current area of investigation \citep{Bakx,Maion}.

Perturbative schemes that go beyond linear order have the interesting consequence that the alignment signal exhibits a B--mode. This B--mode is not expected to arise due to weak lensing itself, at least in a leading approximation \citep[but see][]{waerbeke,cooray,hilbert,hirata_krause,bradshaw}. 
This implies that a detection of such a B--mode which is compatible with predictions of existing alignment models can represent a clean and important validation test of the IA model in question, independently of lensing. It is known from N--body simulations that dark matter halos exhibit a B--mode IA signal compatible with expectations from perturbative models \citep[see e.g.][]{Kurita,Bakx}. In comparison, nowadays, the B--mode signal is mostly utilised as a null check for systematics in the weak gravitational lensing measurements in the literature \citep[see e.g.][and references therein]{asgari}. 

Galaxy clusters are among the largest gravitationally bound objects known in the Universe, and they exhibit strong IA \citep[see e.g.][]{smargon,vanuitert}. We additionally expect galaxy clusters to be relatively clean tracers of their dark matter halo. For these reasons, they offer a useful first opportunity to validate quasi--linear alignment models and a starting point for studying the IA signal of different tracer populations. A validation of the presence of B--modes from IA could point to the need to include such quasi--linear terms in weak lensing analyses of galaxy shapes. In such a scenario, B--modes might play a role as a source of information in alignment bias parameters \citep{pen00}.

In this work, we explore the possibility of detecting an alignment B--mode in galaxy clusters detected in the upcoming LSST survey by using Fourier--space multipoles of the B--mode. In Section \ref{sec:Modelling}, we explain the alignment model we employ in this work, which is based on the Effective Field Theory of Intrinsic Alignments (EFT of IA, see below) as well as the observables considered. In Section \ref{sec:Forecasting} we detail the forecasting method and subsequently Section \ref{sec:Results} shows our baseline forecasting result and several robustness checks. Throughout this paper, we assume a cosmological model in agreement with the Planck 2018 results \citep[][TT,TE,EE+lowE+lensing, marginalised means in Table 1]{planck} with $\Omega_m=0.315,\Omega_b=0.049,h=0.674,n_s=0.965$ and $\sigma_8=0.811$.

\section{Modelling}
\label{sec:Modelling}
On large, quasi--linear scales, the alignment of tracers of the dark matter field can be modelled with perturbation theory. In particular, the familiar linear alignment model from \cite{catelan_ia,hirata_seljak} is simply the leading term in an expansion of the IA signal in terms of the dark matter density perturbation $\delta$, which, when smoothed on large enough scales, can be assumed to be small. In the past decade, the Effective Field Theory of Large Scale Structure \citep[EFT of LSS,][]{baumann, carrasco} paradigm has emerged as a consistent way to treat populations of dark matter tracers on cosmological scales \citep{schmidt}. It was applied to IA in \cite{vlah1} and \cite{Bakx}, thus giving rise to the Effective Field Theory of Intrinsic Alignments (EFT of IA) following earlier work by \cite{TATT,schmitz}. 

Alignment B--modes only appear at next--to--leading order in perturbation theory. Indeed, the linear alignment model does not predict any B--mode signal. At second order in perturbation theory however, B--modes do appear. Specifically, the model we consider for the intrinsic shape perturbation reads 
\begin{equation}
\begin{aligned}
    g_{ij} = b_1^g K_{ij}+ b_{\delta K} \delta K_{ij} + b_{KK}\text{TF}(K^2)_{ij},
\end{aligned}   
    \label{eq:shapeexp}
\end{equation}
where the trace--free operators appearing on the r.h.s. are given by 
\begin{equation}
\begin{aligned}
    K_{ij} = \bigg( \frac{\partial_i \partial_j}{\nabla^2}-\frac{1}{3}\delta_{ij}\bigg)\delta &= \frac{2}{3\Omega_m (aH)^2}(\partial_i\partial_j-\frac{1}{3}\delta_{ij}\nabla^2)\Phi,\\
    \text{TF}(K^2)_{ij} &= K_{ik}K_{kj} - \frac{1}{3}\delta_{ij}K_{kl}K_{kl}, \\
\end{aligned}    
\end{equation}
and all quantities are evaluated at position $\mathbf{x}$ and redshift $z$. Here $\Phi $ is the gravitational potential obeying the Poisson equation $\nabla^2\Phi = \frac{3}{2}\Omega_m (aH)^2 \delta$.

This model contains all the terms needed to compute B--mode auto--correlations at next--to--leading order in the EFT framework\footnote{We do not include the operator $t_{ij}$ at second order. This operator is a pure quadrupole in Fourier space and hence does not produce B--modes for the same reason the tidal field $K_{ij}$ does not. Additionally, B--modes do not receive any contributions from third--order operators in the EFT of IA.}. In the flat--sky (i.e. distant observer) approximation where the line--of--sight is the $x^3$--direction, the components of the shape field we are able to observe are 
\begin{equation}
\begin{aligned}
    \epsilon_1 &= \frac{1}{2}(g_{11}-g_{22}), \\
    \epsilon_2 &= g_{12}.
\end{aligned}    
\end{equation}
These fields are coordinate--dependent, but we can construct E-- and B--mode fields (in Fourier space\footnote{We choose to use the notation $f(\mathbf{k})$ rather than $\tilde{f}(\mathbf{k})$ for the Fourier transform of the quantity $f(\mathbf{x})$ to avoid clutter, i.e. the argument $\mathbf{k}$ implies that a Fourier transform is considered.}) as
\begin{equation}
    \begin{aligned}
        E(\mathbf{k}) &= \epsilon_1(\mathbf{k})\cos 2\phi_{\mathbf{k}} + \epsilon_2(\mathbf{k})\sin 2\phi_{\mathbf{k}}, \\
        B(\mathbf{k}) &= -\epsilon_1(\mathbf{k})\sin 2\phi_{\mathbf{k}} + \epsilon_2(\mathbf{k})\cos 2\phi_{\mathbf{k}},         
    \end{aligned}
\end{equation}
where $\phi_{\mathbf{k}}$ is the angle between $\mathbf{k}$ and the $x^1$--axis. The Fourier--space auto--spectra of these shape fields are defined as 
\begin{equation}\label{eq:spec}
    \begin{aligned}
        \langle E(\mathbf{k})E(\mathbf{k}') \rangle &= (2\pi)^3 \delta^D(\mathbf{k+k'})P_{EE}(\mathbf{k}), \\
        \langle B(\mathbf{k})B(\mathbf{k}') \rangle &= (2\pi)^3 \delta^D(\mathbf{k+k'})P_{BB}(\mathbf{k}).
    \end{aligned}
\end{equation}
The B--mode auto--correlation is the only non--vanishing two--point statistic involving B--modes; the cross--correlation $P_{EB}$ vanishes in a parity--invariant Universe \citep{Biagetti20}, while the cross--correlation with the matter density $P_{mB} = 0$ even without this assumption. Note that the power spectra $P_{EE}(\mathbf{k}),P_{BB}(\mathbf{k})$ depend not only on $k = |\mathbf{k}|$ but also on $\mu = \mathbf{k}_3/k$, since statistical isotropy is broken by the line--of--sight direction $x^3$. 

In the context of weak lensing, one typically considers observables that are integrated along the line--of--sight and are thus functions of the projected separation (or in Fourier space, the length of the perpendicular mode $\ell$). For example, \cite{vlah2} derived expressions for flat--sky angular power spectra $C_{XY}(\ell)$ of IA. Unsurprisingly, if one instead attempts to use full three--dimensional information on the positions of the tracer, higher signal--to--noise can be achieved \citep{Singh_multip,kurita_boss}. All information in the three--dimensional power spectra $P_{XY}$ in Equation \eqref{eq:spec} can be summarised in terms of multipole moments $P^{(\ell)}_{XY}(k)$, defined as
\begin{equation}
    P_{XY}^{(\ell)}(k) = \frac{2\ell+1}{2}\int_{-1}^1 \dd \mu \mathcal{L}_\ell(\mu) P_{XY}(k,\mu),
\label{eq:multipoles}
\end{equation}
where $\mathcal{L}_\ell(\mu)$ is a Legendre polynomial of order $\ell$. 

\begin{figure}
    \centering
    \includegraphics[width=.47\textwidth]{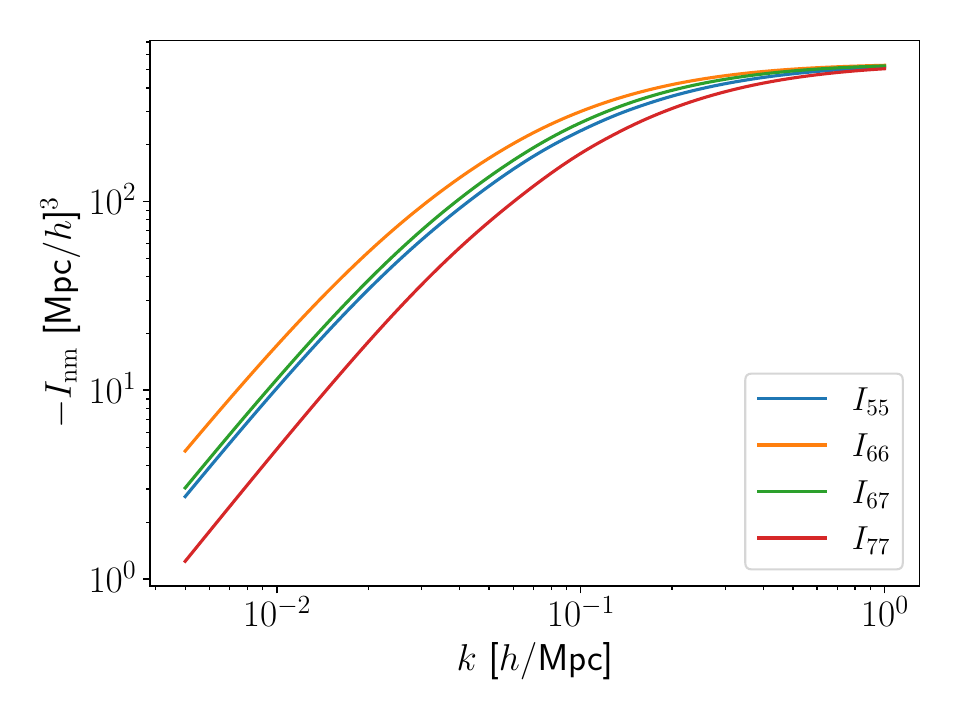}
    \caption{The integrals $I_\mathrm{nm}$, evaluated at redshift zero, that go into the calculation of the B--mode power spectra multipoles, Eq. \eqref{eq:PBB}.}
    \label{fig:Inm}
\end{figure}

Because the 3D shape field is a statistically isotropic rank two tensor\footnote{Throughout this paper, we ignore redshift--space distortions.}, the $\mu$--dependence of the E-- and B--modes is only second order in $\mu$. As such, the associated spectra are at most fourth order in $\mu$, and clearly they should be even in $\mu$. In fact, the B--mode is fully captured by the first two moments $\ell=0,2$. These arguments are based solely on symmetries and hold independently of perturbation theory. Within the perturbative model  of Eq. \eqref{eq:shapeexp}, we obtain the following explicit expressions for the B--mode multipoles: 
\begin{equation}
\begin{aligned}
    P_{BB}^{(0)}(k,z) = &\frac{1}{3}\big[(b_{2,1}^g)^2(2I_{55}+ I_{66})+  2(b_{2,1}^g b_{2,3}^g) (I_{66}-I_{67}) \\
    &+ (b_{2,3}^g)^2 (I_{66}  - 2I_{67} + I_{77})\big],\\
    P_{BB}^{(2)}(k,z) = &\frac{2}{3}\big[(b_{2,1}^g)^2(I_{66}-I_{55}) +  2(b_{2,1}^g b_{2,3}^g)(I_{66}-I_{67}) \\
    &+ (b_{2,3}^g)^2(I_{66} - 2I_{67} + I_{77})\big],
    \label{eq:PBB}
\end{aligned}
\end{equation}
where the bias parameters $b_{2,1}^g$ and $b_{2,3}^g$ are related to those from the basis above via the linear combinations
\begin{equation}
    b_{2,1}^g = \frac{1}{3}b_{KK} + b_{\delta K}; \qquad b_{2,3}^g = -\frac{2}{3}b_{KK} + b_{\delta K}. 
\end{equation}
The linear alignment bias $b_1^g$ does not appear explicitly in the expressions for the B--modes. The expressions for the loop integrals $I_{55}, I_{66}, I_{67}, I_{77}$ as a function of wavenumber $k$ are provided in Appendix \ref{ap:Inm}. We adhere to the notation of \cite{vlah1,Bakx}. These integrals are computed in \textsc{Mathematica} with the FFTLog method from \cite{fftlog}. In Figure \ref{fig:Inm} we show the dependence of these integrals on wavenumber for our fiducial cosmology. Note that both of these expressions scale with redshift as $D(z)^4$, since the loop integrals $I_{\text{nm}}(k,z)$ involve the square of the linear power spectrum.

We emphasise that these relations are valid for any tensorial tracer of large--scale structure, regardless of its precise definition. This includes halos, galaxies, and galaxy clusters. In this framework, modelling the galaxy--halo connection (or the cluster--halo connection, for that matter) is not necessary, which is a major advantage in terms of reducing the number of free parameters of the model. 

We expect the EFT framework to yield valid predictions up to wavenumbers of order the non--linear wavenumber $k_\mathrm{NL}$, which we define here through the relation \citep{kNL},
\begin{equation}
    k_\mathrm{NL}^{-2}(z)=\frac{1}{12\pi^2}\int_0^\infty P_\mathrm{L}(k',z)\dd k'\,.
    \label{eq:knl}
\end{equation}
Note that $k_\mathrm{NL}$ grows with increasing redshift, because gravitational non--linearities grow with time. Put differently, the (RMS) density contrast was smaller in the past, because structure has had less time to grow. From Eq. \eqref{eq:knl} we can deduce that $k_\mathrm{NL}(z) \propto D(z)^{-1}$ where $D(z)$ is the growth factor. In \cite{Bakx} it was demonstrated that the EFT framework describes the B--mode signal for dark matter halos successfully up to $k_\mathrm{max}\simeq0.28$ $h/$Mpc at $z=0$. If we assume that the maximum wavenumber is proportional to $k_\mathrm{NL}$ at all redshifts, we see that $k_{\mathrm{max}}$ also grows as $D(z)^{-1}$ with increasing redshift. Hence, 
\begin{equation}
    k_{\text{max}}(z) = k_{\mathrm{max}}(z_*)\frac{D(z_*)}{D(z)} = (0.28\,h/\text{Mpc})\frac{1}{D(z)}.
\end{equation}
This is the strategy we adopt to determine scale cuts in this work. That is, we assume the range of validity of the perturbative expansion to be the same for clusters as for the dark matter halos hosting them\footnote{It is possible that very massive tracers require the inclusion of higher--derivative terms \citep{hdbias}. This is beyond the scope of this paper, although we may note that the first term of this kind, namely $\nabla^2K_{ij}$, does not contribute to B--modes. Nontrivial corrections would arise from spatial derivatives of second order quantities.}.

In principle, all coefficients appearing in Eq. \eqref{eq:shapeexp} are independent, if no assumptions are made about the origin and evolution of the IA signal. However, several works have suggested that, for dark matter halos, to a good approximation, intrinsic alignment is local in Lagrangian space \citep{Bakx,Maion,Akitsu_coevol}. That is to say, the linear alignment model is a good approximation in the far past, at the initial time when alignments originated. The non--linear terms appearing in Eq. \eqref{eq:shapeexp} then arise due to advection effects displacing the halo from its initial position. This treatment yields the following co--evolution relations among the higher--order bias parameters \citep{Bakx}: 
\begin{equation}
    %\begin{aligned}
        b_{2,1}^\mathrm{g}=(b_1^\mathrm{s}-1)b_1^\mathrm{g},\quad
        b_{2,3}^\mathrm{g}=b_1^\mathrm{s} b_1^\mathrm{g},
    %\end{aligned}
    \label{eq:coevolution}
\end{equation}
where $b_1^s$ is the linear galaxy bias defined through $\delta_g = b_1^s \delta$. 

Strictly speaking, the co--evolution relations have been verified only for halos (rather than clusters). The work of \cite{Akitsu_coevol} applied the quadratic field method from \cite{schmittfull} to measure second order bias parameters with high accuracy. Their measurements showed good agreement with the co--evolution relations in N-body simulations across a halo mass range of $ \sim 10^{12} -  10^{15}M_{\odot}/h$ and for redshifts $0<z<1$. The cluster samples we will consider below have a typical host halo mass of $\sim 10^{14} M_{\odot}/h$. (It is safe to assume that at these masses, every such halo hosts a galaxy cluster -- e.g., \citealt{Kravtsov}.) Moreover, \cite{Shi} found using cluster mock catalogues from \cite{sunayama} that the shape of the dark matter halo traces well the shape of the cluster (as determined by using all the member galaxies). For this reason, we expect that clusters trace the shapes and positions of their host halos very well, and therefore the co--evolution relations hold for clusters too. One caveat in this argument, as is argued in \cite{Shi}, is that the shape of clusters as determined by cluster finders using photometric redshift data is not equal to the true shape of the cluster because of measurement uncertainties in the redshift and missed satellite members. Their work indicates that this reduces the entire alignment signal by a factor of $\sim 2.5$ at separations $r_p \sim 6-70$ Mpc $/h$ irrespective of cluster richness, which includes the scales relevant for this work. Hence the observed cluster alignment also obeys the co--evolution relations, as both sides of Equation \eqref{eq:coevolution} are linear in the shape bias parameters. The reduction in the alignment signal is taken into account in our analysis, because the model we use for estimating $b_1^g$ for clusters is based on observational data rather than estimates from simulations.

\section{Forecasting setup}
\label{sec:Forecasting}

The objective of this analysis is to estimate the signal of B--mode shape--shape correlations of galaxy clusters that is expected to be observed by next generation galaxy surveys. In addition, we are interested in estimating the noise to this observable and quantify the expected significance of the detection of the B--mode signal. Specifically, we will focus on estimating this signal for an experiment setup that simulates the upcoming LSST survey. To this end, we need to estimate the redshift distribution of the observed galaxy clusters and the covariance matrix of the measured signal. In this section, we outline the procedure to do so.

\subsection{Clusters in LSST}

We use the halo model formalism to predict the expected number of observed galaxy clusters. According to this, the cluster count can be estimated by 
\begin{equation} 
    \frac{\dd N}{\dd z}=\Omega_\mathrm{s}\,\frac{\dd^2 V_\mathrm{c}}{\dd z\dd\Omega}\int\dd M \frac{\dd n}{\dd M}\int_{\lambda_\mathrm{min}}^{\lambda_\mathrm{max}}\frac{\dd\lambda}{\lambda}\,p(\ln\lambda|M)\,,
    \label{eq:cluster_counts}
\end{equation}
where $\Omega_\mathrm{s}$ is the survey area, $V_\mathrm{c}$ the comoving volume, $n(M,z)$ the halo mass function and $\lambda$ is the richness of the cluster, with the rightmost integral carried out between the minimum and maximum richness considered. We use the model from \cite{Tinker} for the halo mass function.

The probability density function $p(\ln\lambda|M)$ is the mass--richness relation and quantifies the distribution of cluster richness given its halo mass and redshift. We model this relation following \cite{mass-richness, Murata},
\begin{equation}
    p(\ln\lambda|M)=\frac{1}{\sqrt{2\pi}\sigma_{\ln \lambda|M}}\exp\left(-\frac{\left(\ln\lambda-\langle\ln\lambda\rangle(M)\right)^2}{2\sigma^2_{\ln \lambda|M}}\right)\,.
    \label{eq:mass-richness}
\end{equation}
The distribution above is a log--normal with the mean relation often parameterised by
\begin{equation}
    \langle\ln\lambda\rangle(M)=A+B\ln\left(\frac{M}{M_\mathrm{piv}}\right)\,
    \label{eq:mean_mass-richness}
\end{equation}
where typically it is chosen that $M_\mathrm{piv}=3\times10^{14}$ M$_\odot/h$ and the scatter given by
\begin{equation}
    \sigma_{\ln \lambda|M}=\sigma_0+q\,\ln\left(\frac{M}{M_\mathrm{piv}}\right)\,.
    \label{eq:scatter_mass-richness}
\end{equation}
In both the mean and scatter of the mass--richness relation, a redshift dependence can also be considered, in the form of an additive $\ln(1+z)$ term, multiplied by a factor $q_z$. However, no such term is supported by current data and we do not consider it in this analysis. 

We assume the observed cluster sample to be volume limited and base our forecast on a sample generated from red sequence cluster finders, such as redMaPPer \citep{redmapper}. This is the same cluster finder that was used in the analysis of \cite{Murata} and  we use the best--fit values for the parameters of the mass--richness relation from that work. Specifically, we use $A=3.207$, $B=0.993$, $\sigma_0=0.456$ and $q=-0.169$. In the integration of Equation \eqref{eq:mass-richness} we use $\lambda_\mathrm{min}=20$, $\lambda_\mathrm{max}=100$ and integrate over $10^{12}\le M/[\mathrm{M}_\odot/h]\le 2\times10^{15}$, following the ranges used in \cite{Murata}.

The redshift out to which the cluster sample is volume limited is determined by the depth of the survey and the wavelength at which the observations are made. When the $4000$ $\mathrm{\AA}$  break is redshifted beyond the wavelength of observations, the galaxy red sequence can no longer be detected and that defines the limit of the cluster sample. However, the survey must also be deep enough to detect enough galaxies out to that redshift. A galaxy of $0.2L_*$ luminosity needs to be detected at that redshift for optimum richness estimation of the volume limited cluster catalogue \citep{redmapper}.

The filters for LSST are $u,g,r,i,z$ and $y$, with $y$ being the reddest available filter. The turn--on point for the $z$ filter\footnote{\url{https://noirlab.edu/science/programs/ctio/filters/Dark-Energy-Camera}} is approximately at $8500$ $\mathrm{\AA}$, which means the maximum redshift that the red sequence can be reliably estimated is $z_\mathrm{max}\approx1.1$. The limiting magnitude in the $i$--band is also forecasted to be $i_\mathrm{lim}=25.3$ for the Year 10 LSST data \citep{SRD}. Most recently, the redMaPPer algorithm was applied on data from the Dark Energy Survey Year 1 data release \citep[DESY1,][]{DESY1, redmapper_desy1} using filters up to $z$--band for the determination of the galaxy red sequence.  The limiting magnitude in the $i$--band for DESY1 is $i_\mathrm{lim}\approx22.9$. We therefore assume\footnote{In \cite{redmapper} a galaxy with $L=0.2L_*$ has apparent magnitude $m_*(z)+1.75$ mag, with $m_*(z)$ given by Eq. (6) of that article. This equation holds for $z<0.7$, but an extrapolation to $z=1.1$ gives a limiting magnitude of 24.85 mag. As we will see in Section \ref{sec:Results}, the majority of the SNR is gained at redshift that is lower than 1.1 hence inaccuracies in the magnitude limit estimation are not expected to qualitatively alter our results.}
that survey depth will not be an issue and that our forecasted cluster sample is volume limited out to redshift of $z_\mathrm{lim}=1.1$.

We note here that the fiducial value for $q$ in Eq. (\ref{eq:scatter_mass-richness}) has been chosen as 0 in several other forecasts of LSST clusters \citep[e.g.][]{SRD, Eifler}. These parameters can depend on the cluster finder algorithm and even on the specific setup choices of the algorithm. In this work, we do not assume $q$ to be zero, since our forecast is based on the same cluster finder algorithm and setup as in \cite{Murata}, where $q$ was shown to be significantly non--zero. In Section \ref{sec:Results}, we repeat our analysis with slightly different values for the mass--richness relation parameters and show how these affect our results.

Lastly, we will assume the redshift estimates of the galaxy clusters to be precisely known for the main analysis. This will allow us to study the maximum B--mode information available in such a cluster sample from two--point correlations. Accurate redshifts for these clusters can potentially be obtained by overlapping spectroscopic surveys, which usually target luminous red galaxies, the main population of galaxies in galaxy clusters. However, since the LSST survey will be entirely photometric and cluster redshift estimates will not be as precise, and we address the impact of redshift uncertainty later in Section \ref{sec:photoz}.

\subsection{Bias coefficients}

Using the halo model formalism we are also able to compute the average linear galaxy cluster position bias $b_1^\mathrm{s}$ using the halo mass function $n(M,z)$, the halo bias function $b_\mathrm{h}(M,z)$ and the mass--richness relation. We can write
\begin{equation}
    b_1^\mathrm{s}=\frac{1}{\bar{n}}\int\mathrm{d} M\,\frac{\mathrm{d}n}{\mathrm{d} M} \,b_\mathrm{h}(M,z)\int_{\lambda_\mathrm{min}}^{\lambda_\mathrm{max}}\frac{\dd\lambda}{\lambda}\,p(\ln\lambda|M)\,,
    \label{eq:b1s}
\end{equation}
where $\bar{n}$ is the average number density given by
\begin{equation}
    \bar{n}=\int\mathrm{d} M\,\frac{\mathrm{d}n}{\mathrm{d} M} \int_{\lambda_\mathrm{min}}^{\lambda_\mathrm{max}}\frac{\dd\lambda}{\lambda}\,p(\ln\lambda|M)\,.
    \label{eq:nbar}
\end{equation}
The halo bias function we use is described in \cite{Tinker}.

In a similar way, we can compute the average linear alignment bias $b_1^\mathrm{g}$ by assuming a mass dependence of the amplitude of the linear alignment model, $A_\mathrm{IA}$. To model this, we use measurements from Fortuna et al. (in prep.), who measured the mass dependence of the linear alignment amplitude from several samples of luminous red galaxies and clusters from the SDSS redMaPPer cluster catalogue \citep{redmapper,vanuitert}. This relation had already been shown to hold for dark matter halos \citep{Piras, Kurita}. 
Therefore, we assume a power--law model,
\begin{equation}
    A_\mathrm{IA}(M)=A_\mathrm{IA,0}\left(\frac{M}{M_\mathrm{p}}\right)^{\beta_\mathrm{M}}\,,
    \label{eq:AIA(M)}
\end{equation}
with a pivot mass $M_\mathrm{p}=10^{13.5}$ M$_\odot/h$. The alignment amplitude is related to the linear alignment bias via 
\begin{equation}
    b_1^\mathrm{g}(M,z)=-2A_\mathrm{IA}(M)\frac{C_1\rho_\mathrm{crit}\Omega_m}{D(z)}\,,
    \label{eq:b1g_AIA}
\end{equation}
where $\rho_\mathrm{crit}$ is the critical density of the universe and $C_1=5\times10^{-14}$ M$_\odot^{-1}h^{-2}$Mpc$^{-3}$, a normalisation constant typically used in intrinsic alignment measurements, first determined by \cite{C1}. Following Equation \eqref{eq:b1g_AIA}, we can estimate
\begin{equation}
    b_1^\mathrm{g} = \frac{1}{\bar{n}}\int\mathrm{d} M\,\frac{\mathrm{d}n}{\mathrm{d} M} \,b_1^\mathrm{g}(M,z)\int_{\lambda_\mathrm{min}}^{\lambda_\mathrm{max}}\frac{\dd\lambda}{\lambda}\,p(\ln\lambda|M)\,.
    \label{eq:b1g}
\end{equation}

We use the best--fit values from Fortuna et al. (in prep.) and have $A_
\mathrm{IA,0}=5.74$ and $\beta_M=0.44$. The amplitude, $A_\mathrm{IA,0}$, was determined using redMaPPer galaxy clusters to pinpoint the IA behaviour on high masses. The shape of these galaxy clusters are obtained from positions of observed satellite members, weighted by their membership probability. As noted in Section \ref{sec:Modelling}, these shapes have been shown to suffer from projection effects and are expected to have lower alignment amplitude compared to alignments of dark matter halos \citep{Shi}. However, we don't need to correct for this since our forecasting setup is based on a redMaPPer--like galaxy cluster sample. Thus, we use the value of $A_\mathrm{IA,0}$ as is, where the projection effects are implicitly already taken into account.

\subsection{Covariance estimation}

We model the expected error on the measured B--mode IA signal with a Gaussian covariance following \cite{covariance, Taruya}. The covariance is given by
\begin{equation}
    \begin{aligned}
        \mathrm{Cov}^{\ell\ell'}(k_n)=&\frac{(2\ell+1)(2\ell'+1)}{V_n\, V_s}\\
        &\int_{-1}^1 \dd \mu \,\mathcal{L}_\ell(\mu)\mathcal{L}_{\ell'}(\mu)\left(P_\mathrm{BB}(k_n,\mu)+\frac{\sigma_\epsilon^2}{\bar{n}}\right)^2\,,
        \label{eq:covariance}
    \end{aligned}
\end{equation}
where $\sigma_\epsilon$ is the RMS ellipticity of the galaxy cluster sample, $n$ is the number density, $V_s$ is the volume of the survey and $V_n$ the volume element of the thin shell in Fourier space, given by 
\begin{equation}
    V_n = \frac{4\pi^2 k_n^2\dd k_n}{(2\pi)^3}\,.
\end{equation}
The volume element above is computed in bins of $k$ with width $\dd k_n$ and central value $k_n$. 
At each combination of $\ell,\ell'$, the covariance matrix has $N_k\times N_k$ dimensions, where $N_k$ is the number of $k$--bins used in the analysis.
The SNR is then given by
\begin{equation}
    \left(\frac{\mathrm{S}}{\mathrm{N}}\right)^2=\sum_{\ell,\ell'}\left(P_{BB}^{(\ell)}(k_n)\right)^\mathrm{T} \left(\mathrm{Cov}^{\ell\ell'}(k_n)\right)^{-1} P_{BB}^{(\ell')}(k_n)\,,
\end{equation}
where the sum is carried over $\ell,\ell'\in\{0,2\}$.

The covariance above ignores the connected non--Gaussian terms and super--sample covariance terms \citep{SSC, Kurita}. The mean redshift of our forecasted galaxy cluster sample is $\bar{z}=0.676$, which gives $k_\mathrm{max}\approx0.4$ $h/$Mpc. In \citet[Figure 9]{Kurita} it was shown that the Gaussian covariance overestimates the signal--to--noise ratio by approximately 25\% at $k_\mathrm{max}\approx0.4$ $h/$Mpc when considering the halo number counts -- E--mode power spectra, $P_\mathrm{hE}$. We expect this effect to be less dramatic when considering B--mode auto--power spectra, because these are primarily shape--noise dominated. As will be clear in Section \ref{sec:Results}, our results do not change qualitatively even if the SNR is reduced by 25\%. Hence, we limit our analysis to modelling the Gaussian covariance only.

One additional assumption that goes into this covariance is that the shape noise statistics are purely Poissonian, meaning the only contribution to the Gaussian noise is the Poissonian shape noise, $\sigma_\epsilon^2/\bar{n}$. An additional contribution arises from the non--linear evolution of IA \citep{TATT}. This non--Poissonian shape noise is measured in \cite{Kurita} to be 5--10\% of the Poissonian contribution. In Section \ref{sec:Results}, we repeat our analysis with a shape noise that is increased by 10\% and show the effect on the overall SNR is small.

\section{Results}
\label{sec:Results}

In our fiducial setup, we consider a galaxy cluster sample obtained via a redMaPPer--like algorithm applied on LSST Y10 data, the final data products from 10 years of observations. The sample is assumed to be volume limited. The minimum redshift is assumed to be 0.2, similarly to what was found in \cite{redmapper_desy1}, while the maximum redshift is  1.1 (see Section \ref{sec:Forecasting}). The redshift distribution is obtained via Eq. \eqref{eq:cluster_counts} and the total number of clusters forecasted is approximately 124,000 with a mean redshift of $\bar{z}=0.676$ and a total number density of $\bar{n}=4.44\times10^{-7}$ [Mpc$/h$]$^{-3}$, corresponding to a characteristic scale of 130 Mpc$/h$. We assume the cluster redshifts to have no uncertainty and discuss the impact of redshift uncertainty in Section \ref{sec:photoz}.

We compute the B--mode theory power spectra in 117 logarithmically spaced bins of $k$, with central values $k_n$ from $0.005$ $h/$Mpc up to $1.0$ $h/$Mpc. The width of the base--10 logarithm of each bin is given by 0.02. The power spectra are computed at the mean redshift of the survey. We assume a survey with $f_\mathrm{sky}=0.43$ sky coverage fraction \citep{LSST}. We also assume an ellipticity RMS dispersion of $\sigma_\epsilon=0.098$, based on what has been measured from existing redMaPPer cluster data \citep{Vedder}.

\begin{figure}
    \centering
    \includegraphics[width=0.47\textwidth]{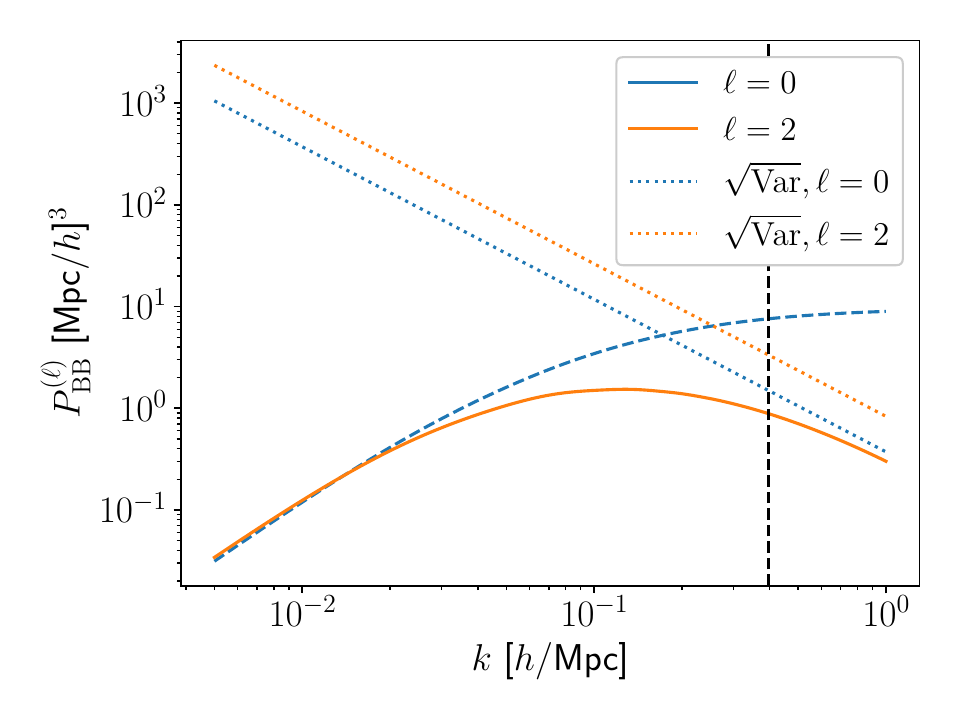}
    \caption{The signal (solid/dashed lines, computed at the cluster sample mean redshift) and noise (dotted lines, square--root of the variance) of the forecasted B--mode power spectra (monopole and quadrupole, $\ell=0$ and $\ell=2$, respectively). Solid lines represent positive signal and dashed ones negative. The vertical dashed black line shows the $k_\mathrm{max}$ at the redshift considered in our fiducial analysis when calculating the SNR.}
    \label{fig:PBB_signal_noise}
\end{figure}

The expected IA signal of the galaxy cluster sample together with the forecasted error on the signal  are shown in Figure \ref{fig:PBB_signal_noise}. On large scales, the B--mode signal is several orders of magnitude lower than the expected level of noise. On smaller scales, the signal increases, especially on the monopole, which drives all the SNR. Around $k\sim0.17$ $h/$Mpc the monopole signal is equal to the square--root of the variance. As the $k_\mathrm{max}$ increases, the SNR is also increasing. We note, however, that this is unlikely for a more realistic covariance and the actual SNR is expected to plateau around $k_\mathrm{max}\approx0.4$ $h/$Mpc \citep{Kurita}. This is also the maximum wavenumber we consider in our fiducial analysis.

The resulting SNR for our fiducial analysis is equal to 11.79. This is much higher than the typical threshold for detection of 5 and suggests that B--mode correlations from galaxy cluster shape--shape 3D power spectra are going to be detectable with high significance in a survey like LSST. 

\begin{figure}
    \centering
    \includegraphics[width=0.47\textwidth]{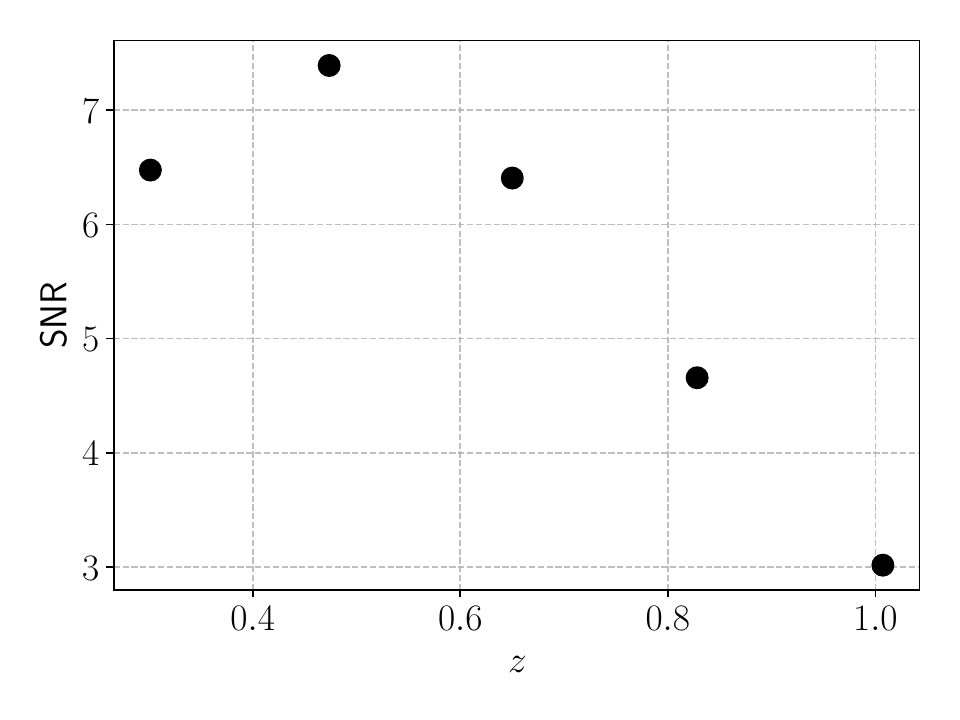}
    \caption{The dependence of the SNR of the forecasted B--modes of LSST cluster alignments as a function of redshift. The points show the expected SNR in each redshift bin assuming 5 equidistant bins between $0.2<z<1.1$. The $x$--axis shows the mean redshift of each bin.}
    \label{fig:SNR_z}
\end{figure}

So far, we have assumed that the signal measured from shape--shape correlations of the cluster sample can be described by computing the theory power spectra at the mean redshfit of the sample. This is a good approximation when the redshift bin has a small width but might not be accurate over larger redshift baselines. However, the result over the full redshift range is still indicative of the expected SNR, and we keep our fiducial setup with this caveat.

Motivated by this, it is interesting to explore the expected SNR in a smaller redshift bin, which will also reveal the redshift that drives the constraining power. Therefore, we divide the redshift distribution into 5 equidistant redshift bins and repeat the analysis in each bin. The lowest redshift bin is the most dense in cluster population, with a characteristic cluster separation of 70 Mpc$/h$. In Figure \ref{fig:SNR_z} we show the forecasted SNR evaluated for each redshift bin and plotted at the mean redshift of each bin. We see that the intermediate redshift bins around 0.45 have a higher SNR. This is important to note since the analysis choices were motivated by cluster samples at low redshift, below 0.35. While the true SNR might be different due to extrapolation of these choices to higher redshift, it is not expected to differ by a very significant amount, since the low redshift clusters contain a large portion of the constraining power. In addition, already at the first redshift bin with mean redshift of 0.3, the forecasted SNR is significant.

It is also useful to note that the overall SNR from binning the sample is higher than the SNR obtained when the sample is analysed as a whole. This can be understood in two ways: a larger redshift baseline means that the total volume of the sample increases and can contribute negatively to the overall SNR due to the increased noise that comes with it. In addition, a large redshift bin includes correlation between shapes of very distant galaxies, which are expected to be very low. Therefore, it is clear that a binning strategy is preferred in increasing the overall SNR of the measurement, and the one presented in Figure \ref{fig:SNR_z} is only one such possible strategy. However, optimising the binning is more relevant when measuring this signal in real data and beyond the scope of this work.

\subsection{Testing the SNR robustness}
\label{sec:robustness}

As noted above, the forecast is based on an extrapolation of existing data to the depth and redshift of LSST. Moreover, there is an inherent uncertainty on the fiducial values we have chosen, which might significantly impact our forecasted SNR. In this section, we vary these analysis choices within their expected errors from observations and check how the forecasted SNR is affected and whether it remains high enough for a conclusive detection.

We decide to vary these parameters individually, since it is difficult to assess the correlation between them. This might be possible for the same class of parameters, such as the linear alignment bias mass scaling or the mass--richness relation. Nevertheless, we choose to vary them independently for simplicity and note that this is pessimistic since correlations in the mass--richness relation restrict the allowed values of these parameters in a way that improves the SNR \citep[see e.g.][Figure 6]{Murata}.

We vary $A_\mathrm{IA,0},\beta_M,$ by 2 times the error on their measurement in each direction, with $\sigma_{A_\mathrm{IA,0}}=0.32$ and $\sigma_{\beta_M}=0.04$. We also vary the mass--richness relation parameters in the same way, where $\sigma_A=+0.044/-0.046$, $\sigma_B=+0.041/-0.055$, $\sigma_{\sigma_0}=+0.047/-0.039$ and $\sigma_q=+0.035/-0.026$ (meaning we vary them differently in the positive and negative direction). We also run our fiducial analysis with a higher assumed $\sigma_\epsilon^2$ by 10\% and another time without evolving $k_\mathrm{max}$ with redshift but fixing it to the value at $z=0$. 

\begin{figure}
    \centering
    \includegraphics[width=0.47\textwidth]{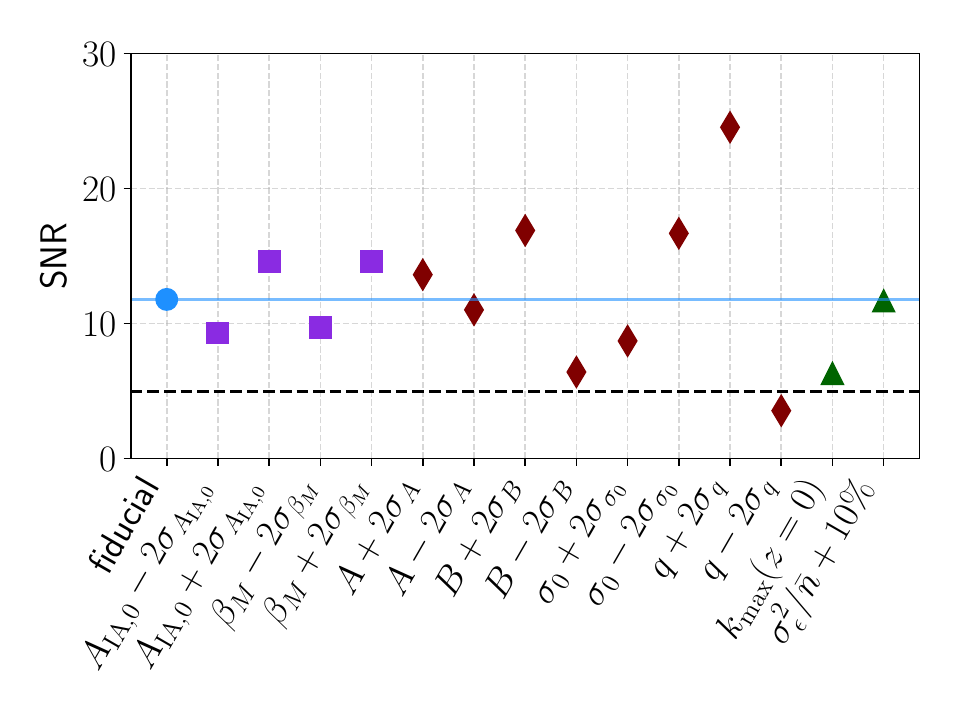}
    \caption{Signal--to--noise ratio of the forecasted B--mode power spectra from intrinsic alignments of galaxy clusters in LSST, for a number of different analysis choices, grouped by colour and symbol. Several setup parameters are allowed to vary individually within twice the value of their standard error. The last two points correspond to fixing the maximum $k$ to redshift 0 and to increasing the $\sigma_\epsilon$ in the covariance by 10\%. The horizontal dashed line shows the threshold for detection, $\mathrm{SNR}=5$. The horizontal blue line shows the $\mathrm{SNR}$ obtained for our fiducial analysis choices.}
    \label{fig:SNR}
\end{figure}

The results from these tests are presented in Figure \ref{fig:SNR}. We see that for most of these tests the SNR is significant, above the threshold of 5 for detection. However, it is clear that the mass--richness relation parameters have the largest impact in the signal, with $B$ and $q$ being the most relevant. These two parameters are also highly negatively correlated \citep{Murata}, where a low value for $B$ is preferred only for a value of $q$ that is closer to zero (note that $q$ is negative, so $q+2\sigma_q$ is closer to zero) and vice versa. Therefore, the negative impact of one of these in the SNR might be alleviated by the expected shift of the other in the opposite direction.

It is also clear that assuming a low $k_\mathrm{max}$ has a drastic effect on the SNR, as is evident on Figure \ref{fig:PBB_signal_noise}. This highlights the importance of developing accurate models of the intrinsic alignment power spectra down to small scales in measuring the B--modes with a high SNR.

\subsection{Forecast for existing cluster catalogues}
\label{sec:stage-III}

Since many cluster catalogues have already been produced from previous galaxy surveys, it is interesting to ask whether a B--mode signal could be observed in shape correlations of these clusters, assuming available precise redshift information. In particular, the redMaPPer cluster catalogue from the Sloan Digital Sky Survey \citep[SDSS,][]{SDSS, redmapper} has been used to produce the highest signal--to--noise measurements of galaxy cluster alignments in data to date \citep{vanuitert}. A similar, deeper catalogue on a much smaller area has been generated using data from DESY1 \citep{redmapper_desy1}. Here, we forecast the B--mode SNR from these two cluster catalogues. 

We choose to model the characteristics of these catalogues with the same methodology used for the LSST cluster sample, instead of using the released data. We have checked that our methodology reproduces the redshift distributions of these catalogues. We leave our fiducial setup unchanged, since the mass--richness relation and alignment bias mass scaling were determined on the SDSS redMaPPer cluster catalogue itself. 

For the SDSS redMaPPer catalogue, we use a survey area of $\Omega_\mathrm{s}=10,400$ deg$^2$ and a redshift range of $0.1<z<0.35$, where the sample is shown to be volume limited. We forecast a SNR $=5.23$, which is slightly above the threshold of $5$ for a detection. Repeating this analysis for the DESY1 sample, we use $\Omega_\mathrm{s}=1,500$ deg$^2$ and $0.2<z<0.7$, and we forecast SNR $=3.28$. This result suggests that a marginal detection might already be within reach with current galaxy cluster catalogues. However, the current spectroscopic surveys do not cover a large enough portion of these catalogues. The redshift information comes from photometry with inherent uncertainty in its value. We discuss this in the section below.

\subsection{Impact of redshift uncertainty}
\label{sec:photoz}

So far in our analysis we have assumed perfect knowledge of the redshift of the galaxy clusters in the forecasted survey. This was done with the aim of exploring the signal itself and understanding the potential for measuring the B--mode correlations in the case where accurate, spectroscopic redshifts can be obtained. In reality, galaxy cluster catalogues in next-generation photometric surveys will rely on photometric redshift estimates. While the accuracy of these redshifts is generally higher than what is typically achievable by photometric samples, they will greatly impact the measurement of the power spectra multipoles. In this section, we take a look at this impact to the B--mode monopole which, as seen in Figure \ref{fig:PBB_signal_noise}, carries most of the signal.

To include the redshift uncertainty in our model, we follow \cite{sigma_chi} and assume that the redshift estimates follow a Gaussian distribution centred on the true redshift with a scatter of $\sigma_z$. For redMaPPer clusters, this uncertainty was quantified to be $\sigma_z/(1+z)\approx0.006$, which we assume here \citep{redmapper_desy1}. This uncertainty propagates into an uncertainty of the comoving radial distance $\sigma_\chi=\sigma_z\mathrm{d}\chi/\mathrm{d}z$ which, in turn, results in a dampening of the power spectra by a factor of $\mathrm{exp(-\mu^2k^2\sigma_\chi^2)}$. We include this factor in the calculations of multipoles, Eq. \eqref{eq:multipoles}.

\begin{figure}
    \centering
    \includegraphics[width=0.47\textwidth]{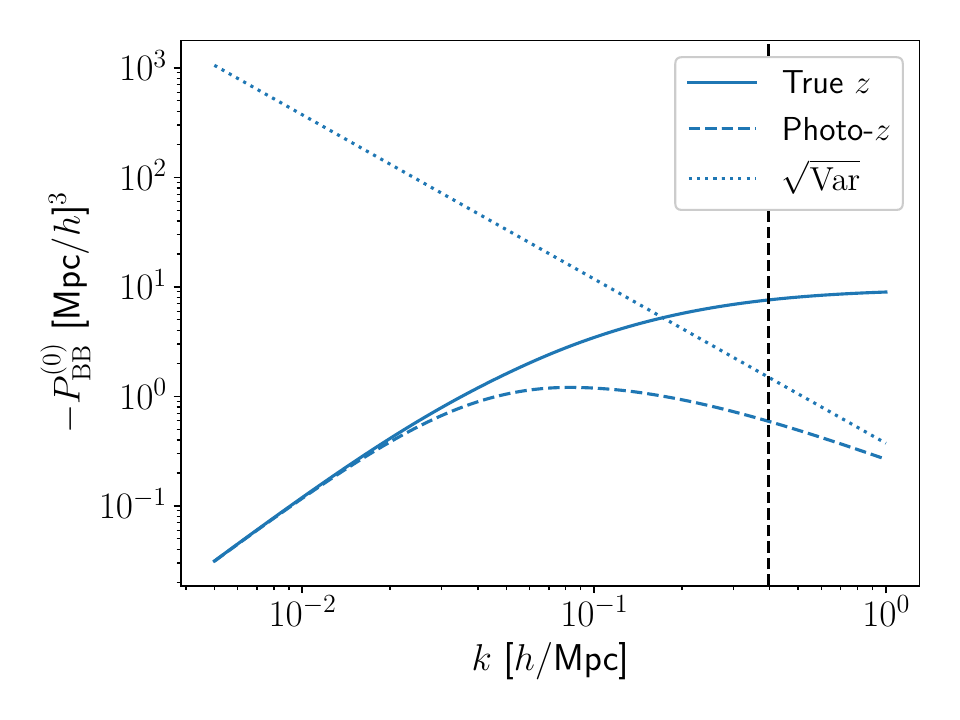}
    \caption{The impact of redshift uncertainty on the predicted B--mode power spectrum monopole. The solid line shows the signal for a sample without any redshift uncertainty while the dashed line shows it for a sample with a redMaPPer--like photometric redshift uncertainty. The signal is computed at the mean redshift of the sample and the dotted line shows the square–root of the variance. The vertical dashed black line shows the maximum wavenumber considered when computing SNR.}
    \label{fig:photoz}
\end{figure}

We show the effect of the redshift uncertainty in Figure \ref{fig:photoz}. The photometric redshifts wash out the 3D information normally accessible by the power spectrum multipoles, dampening the signal significantly. The resulting SNR is also significantly lower, at a value of 2 for our fiducial analysis. However, by following the same redshift binning described in the previous section and summing the SNR at each redshift bin we forecast a total $\mathrm{SNR}=4.93$. This means the B--mode signal from galaxy cluster intrinsic alignments might still be observable in a LSST photometric cluster catalogue, especially if a more optimal binning strategy is possible.

\section{Discussion}
\label{sec:Discussion}

In this work, we forecast the detection of IA B--modes for a sample of galaxy clusters constructed with the redMaPPer algorithm applied on data from the upcoming LSST survey after 10 years of observations. We use existing measurements of the mass--richness relation and mass scaling of the linear alignment bias on similar existing samples to inform our forecast. Assuming perfect knowledge of cluster redshifts, we predict the signal and covariance of the B--mode shape--shape 3D power spectra for such a sample, arising from next--to--leading--order contributions in the EFT model for intrinsic alignments. 

We forecast a high expected SNR of 11.79. We assess the change in SNR when varying the fiducial parameters and analysis choices of the forecast within their observed error and conclude that the SNR does not change significantly as to prohibit a detection. Therefore, we conclude that B--modes from intrinsic alignments of galaxy clusters should be detectable in LSST Y10 data. When including redshift uncertainty from photometric redshifts in the modelling, we find a SNR of 4.9, suggesting a marginal detection is a possibility even without perfect knowledge of cluster redshifts.

Throughout this work, we assume that the shapes of the forecasted clusters are measured using the positions of observed satellite members. Another option is to trace the cluster shape using the shape of the brightest central galaxy (BCG). The alignment amplitude using the BCG shape has been shown to be lower than using the satellite positions, due to the mis--alignment angle of the BCG and the cluster halo \citep{okumura,Shi}. However, the expected SNR is large enough that a BCG B--mode detection might still be worth exploring in the future.

It is interesting to consider whether the B--mode signal will be detectable in earlier data releases from LSST, for example, the Y1 data. These data will cover the same area on the sky as the Y10 data but to a different limiting depth, with forecasts showing $i_\mathrm{lim}=24.1$ \citep{SRD}. This magnitude limit is likely not deep enough to allow the detection of clusters out to $z\sim1.1$, but it is higher than the $i_\mathrm{lim}$ in DESY1 data. This means the LSST Y1 cluster sample will be volume limited between $0.2<z<0.7$ as well, and Figure \ref{fig:SNR_z} indicates the majority of the SNR is gained at those redshifts. Imposing this redshift limit, we forecast that for the LSST Y1 sample the SNR is 11.3, already high enough to be measured and close to the Y10 value. This result stems from the fact that clusters between $0.7<z<1.1$ do not contribute to the SNR significantly. 

The forecast carried out in this work has important implications for cosmic shear surveys that measure shape--shape correlations. Typically, B--modes are assumed to be zero in such measurements, since weak gravitational lensing does not produce them and the statistical power of existing surveys is not high enough to reveal a B--mode signal from the many sources that faintly produce it. However, here we show that, at least for galaxy clusters, such a signal will possibly be detectable in stage--IV cosmic shear surveys, even in the case of photometric redshift estimates. Note also that, while the model studied here is valid up to a maximum scale $k_\mathrm{max}$, B--mode correlations are expected to be non-zero even beyond that limit, at scales typically probed by cosmic shear analyses \citep[see e.g., Figure B1 of][]{Kurita}.

Nonetheless, it is difficult to estimate the expected B--mode signal from intrinsic alignments in future cosmic shear data due to the difficulty in understanding the way the signal manifests in galaxies. The co--evolution relations and linear alignment bias mass scaling that was assumed in this work has been shown to hold for dark matter halo masses between $10^{12}$--$10^{15}$ M$_\odot/h$ and redshift up to $\sim1$ and $\sim0.5$, respectively. Cosmic shear samples typically extend far beyond these limits, especially for upcoming stage--IV surveys. Moreover, while galaxy clusters are "clean" tracers of dark matter halos and these relations are true for them, it is harder to understand how these relations transform for the galaxy samples that are typically employed in cosmic shear analyses. The selection effects for cosmic shear samples need to be carefully taken into account in the halo model formalism and the mis--alignment of galaxies with their host halo must be accurately modelled for such a forecast. 

While forecasting for cosmic shear samples is beyond the scope of this work, it is clear that the B--mode signal of cosmic shear measurements for stage--IV surveys might not be a clean test of systematic errors in the weak gravitational lensing measurements. On the contrary, it is likely a source of extra information that could be leveraged and might contribute to tightening the constraining power of such surveys. For this, the alignment signal needs to be accurately modelled down to quasi--linear scales. However, there exist other sources of B--mode correlations that might also contribute, such as the clustering of source galaxies \citep{source_clustering} or the lensing bias \citep{hirata_krause}. A careful analysis of the relative amplitudes of the various sources of B--mode shape--shape correlations is, therefore, necessary in order to be able to confidently model the measured B--mode signal. We leave this question for future work.

\begin{acknowledgments}
The authors would like to thank Dr. Michel Aguena and Dr. Eli Rykoff for useful discussions on the forecast of the cluster sample. This publication is part of the project ``A rising tide: Galaxy intrinsic alignments as a new probe of cosmology and galaxy evolution'' (with project number VI.Vidi.203.011) of the Talent programme Vidi which is (partly) financed by the Dutch Research Council (NWO). This work is also part of the Delta ITP consortium, a program of the Netherlands Organisation for Scientific Research (NWO) that is funded by the Dutch Ministry of Education, Culture and Science (OCW). We thank Henk Hoekstra and Maria Cristina Fortuna for sharing preliminary results from their manuscript in preparation with us, and Henk Hoekstra for useful discussions.
\end{acknowledgments}

\appendix

\section{Next--to--leading Order Perturbative Contributions to the B--mode Signal }
\label{ap:Inm}

The relevant integrals for B--mode auto--correlations at one--loop order in the EFT of IA are given by
\begin{equation}
\begin{aligned}
    I_{55}(k,z) &= -\int_\mathbf{q} \frac{(\mathbf{k}\cdot\mathbf{q}-q^2)^2(k^2-2\mathbf{k}\cdot\mathbf{q})^2((\mathbf{k}\cdot\mathbf{q})^2-k^2q^2)}{4k^4q^4|\mathbf{k-q}|^4}P_L(q,z)P_L(|\mathbf{k-q}|,z);\\
    I_{66}(k,z) &= \int_\mathbf{q} \frac{(\mathbf{k}\cdot\mathbf{q}-q^2)^2((\mathbf{k}\cdot\mathbf{q})^2-k^2q^2)}{4k^4q^4|\mathbf{k-q}|^4}P_L(q,z)P_L(|\mathbf{k-q}|,z);\\
    I_{67}(k,z) &= -\int_\mathbf{q} \frac{(\mathbf{k}\cdot\mathbf{q}-q^2)(k^2+2q^2-2\mathbf{k}\cdot\mathbf{q})((\mathbf{k}\cdot\mathbf{q})^2-k^2q^2)}{8k^4q^4|\mathbf{k-q}|^4}P_L(q,z)P_L(|\mathbf{k-q}|,z);\\
    I_{77}(k,z) &= \int_\mathbf{q} \frac{(k^2+2q^2-2\mathbf{k}\cdot\mathbf{q})^2((\mathbf{k}\cdot\mathbf{q})^2-k^2q^2)}{16k^4q^4|\mathbf{k-q}|^4}P_L(q,z)P_L(|\mathbf{k-q}|,z);
\end{aligned}
\end{equation}
where the shorthand notation 
\begin{equation}
    \int_\mathbf{q} = \int \frac{\dd ^3 \mathbf{q}}{(2\pi)^3}
\end{equation}
is used for the integration measure in Fourier space. The integrals all asymptote to the same positive, scale--independent constant in the $k \to 0$ limit, and this constant is subtracted before computing the signal for a given set of bias parameters (cf. Figure \ref{fig:Inm}). 

\bibliographystyle{mnras}
\bibliography{references}

\end{document}